\documentclass[a4paper,aps,prl,showpacs,twocolumn,superscriptaddress]{revtex4}
\usepackage{graphicx,t1enc}

\begin{document}

\title{\hspace{70pt}Periodically Varying Externally Imposed \newline
\hspace{-5pt}Environmental Effects on Population Dynamics}

\author{M. Ballard}
\affiliation{Consortium of the Americas for Interdisciplinary
Science and Department of Physics and Astronomy, University of New
Mexico, Albuquerque, NM 87131, U.S.A.}

\author{V. M. Kenkre}
\affiliation{Consortium of the Americas for Interdisciplinary
Science and Department of Physics and Astronomy, University of New
Mexico, Albuquerque, NM 87131, U.S.A.}

\author{M. N. Kuperman}
\affiliation{Consortium of the Americas for Interdisciplinary
Science and Department of Physics and Astronomy, University of New
Mexico, Albuquerque, NM 87131, U.S.A.} \affiliation{Centro
At{\'o}mico Bariloche and Instituto Balseiro, 8400 S. C. de
Bariloche, Argentina}

\begin{abstract}
Effects of externally imposed periodic changes in the environment
on population dynamics are studied with the help of a simple
model. The environmental changes are represented by the temporal
and spatial dependence of the competition terms in a standard
equation of evolution. Possible applications of the analysis are
on the one hand to bacteria in Petri dishes and on the other to
rodents in the context of the spread of the Hantavirus epidemic.
The analysis shows that spatio-temporal structures emerge, with
interesting features which depend on the interplay of separately
controllable aspects of the externally imposed environmental
changes.
\end{abstract}

\pacs{05.10.-a, 87.23.Cc} \maketitle

\vspace{1cm}

\section{ \ Introduction }

The mathematical study of population dynamics has been a subject
of great interest in recent years, with application widely spread
among different fields. An example is the description of emerging
patterns in bacterial colonies
\cite{jaco,waki,lin,nels,lito,thesis,kk}. Other studies describe
the behavior of the populations of superior organisms such as
insects and rodents \cite{chule,ale,oku,murr}.

If, in basic models for the description of the evolution of a
given population, we focus attention on reproduction, competition
for resources, and diffusion, the Fisher equation \cite{fish,murr}
appears to be a useful mathematical tool:
\begin{equation}
\frac{\partial u\left( x,t\right) }{\partial t}=D\frac{\partial
^{2}u\left( x,t\right) }{\partial x^{2}}+au\left( x,t\right)
-bu^{2}\left( x,t\right) . \label{originaleq}
\end{equation}
Diffusion with coefficient $D$ is considered here as well as the
growth of the population at rate $a$ and a competition process
weighted by $b$, also containing environmental features. In the
present paper, we consider effects of spatio-temporal dynamics in
the nonlinear term as a representation of externally imposed or
seasonal environmental variations. We thus take $b$ in Eq.
\ref{originaleq} to be space and time dependent, $b(x,t)$.
Environmental variations could alternatively be considered as
affecting $a$ (such that $a$ is dependent on time and space rather
than $b$) as has been done in some earlier studies \cite{lin,kk}.
However, we restrict our attention to a constant $a$ and varying
$b(x,t)$ because this allows us to define a Fisher velocity and
size in terms of a constant $a$. Following \cite{kk} we consider a
bounded region which will be referred to as a $bubble$ of
favorable conditions suitable for the survival of a given species.
Outside of this region the environmental conditions are so harsh
that there is no possibility of survival.

Associated with the Fisher equation \cite{fish} there is a natural
velocity and a natural length, $2\sqrt{Da}$ and $\pi \sqrt{D/a}$,
respectively. The former is known as the Fisher velocity and is
the velocity at which fronts tend to travel after adjustments from
most initial conditions \cite{murr}. Thus, an arbitrary shape of
the initial population with compact support will have fronts that
move, after initial transients, with the Fisher velocity. It is an
important quantity in experiments on bacteria with moving masks
\cite{lin} as it represents the minimum mask velocity at which
bacteria tend to extinction, rather than being able to follow the
bubble through the combined effect of growth and diffusion. The
Fisher length $\pi \sqrt{D/a}$ arises \cite{kk,skellam} in the
bacterial context as the minimum length of the mask below which
the steady state population of the bacteria vanishes as the
bacteria diffuse out of the masked area quicker than they can grow
to any finite saturation value. The Fisher length is thus the
`diffusion length' within the growth time (reciprocal of the
growth rate $a$).

Consideration of these two quantities suggests that it is natural
to envisage two observations involving externally imposed periodic
variation of the environment. The first is one in which
experiments of the kind reported in ref. \cite{lin} are carried
out with the velocity varying periodically, in other words with
the bubble repeatedly returning to its original position. This is
the \emph{oscillating} bubble case. The second experiment is one
in which the bubble extent is made to vary from a large size to
one below the Fisher length (i.e. the extinction size). The
objective of the latter ($breathing$ bubble) experiment is to
analyze extinction tendencies under periodic variations of the
bubble size. We analyze these two hypothetical experiments in this
paper through numerical calculations based on a Crank-Nicholson
scheme.

\section{Traveling bubble}
\begin{figure}
\centering
\resizebox{\columnwidth}{!}{\includegraphics{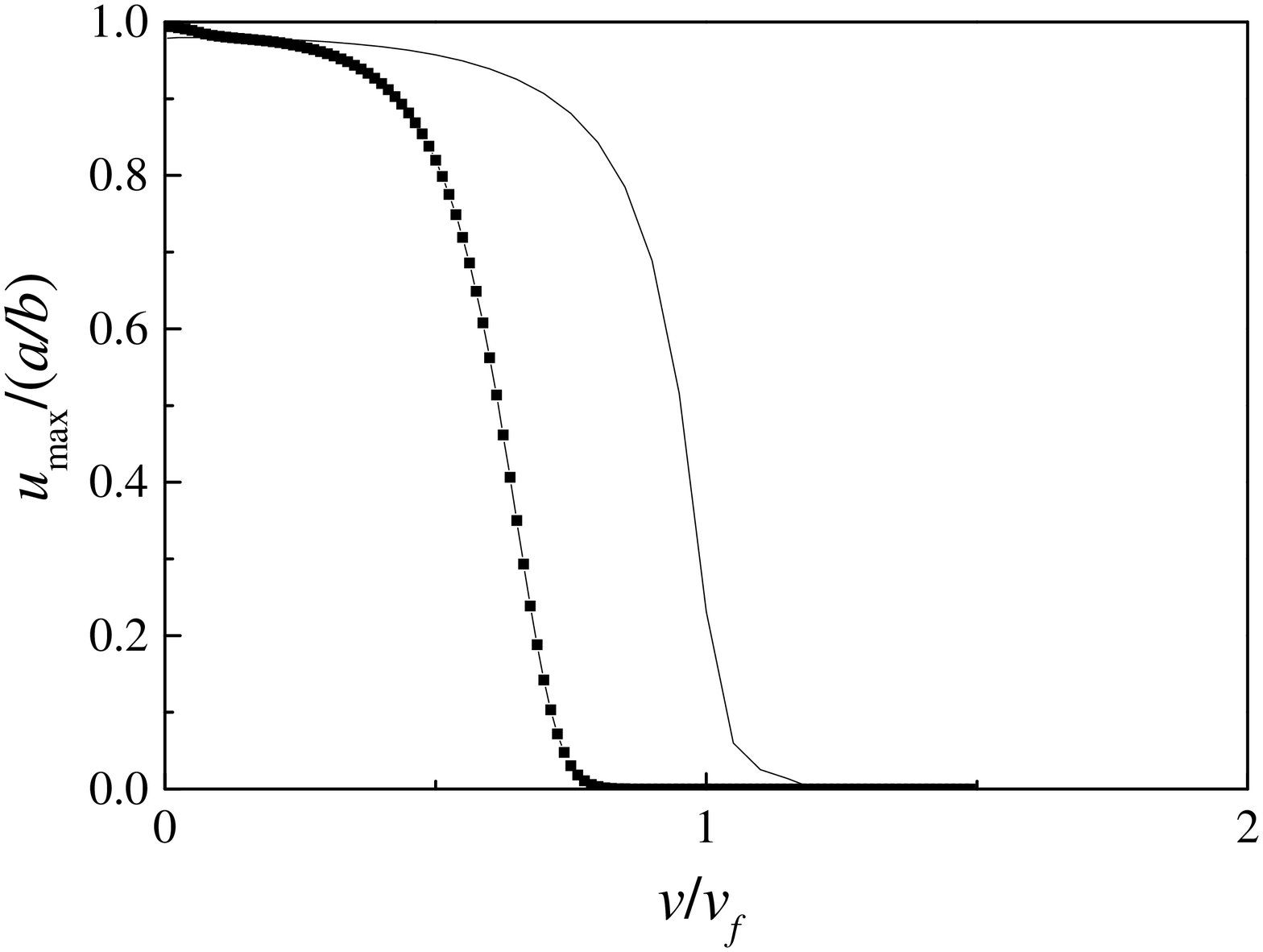}}
\caption{The bold curve depicts $u_{max}$ as the bubble moves
adiabatically; acceleration is affectively zero. The remaining
curve displays $u_{max}$ for a bubble accelerating linearly at
$.1a^2W_c$.} \label{acceleratingbubble}
\end{figure}

Solutions of the Fisher equation for bounded domains,
corresponding to static bubbles, have been already discussed in
\cite{kk} and references therein. A natural change, to be
considered as a first approximation to the problem addressed here,
is to allow a translation in the bubble. We have performed a brief
analysis of situations where the velocity is constant and the
system's behavior corresponds to that of a moving front, as well
as others where the velocity of the bubble changes continuously.
For the former case, and as mentioned in \cite{murr}, we recover a
critical velocity beyond which the front will no longer survive,
see bold curve of \ref{acceleratingbubble}. As $b \rightarrow
\infty$ at the boundaries of the bubble, the curves will abruptly
go to zero, rather than asymptotically as above. The analysis of
this situation can be done by making a change of variables on Eq.
(\ref{originaleq}). If we set $\xi=x-ct$ we obtain the following
equation which is valid in the moving frame accompanying the
bubble.
\begin{equation} D\frac{\partial ^{2}u\left(\xi\right) }{\partial \xi^{2}}
+au\left(\xi\right) -bu^{2}\left(\xi\right) +c \frac{\partial
u\left(\xi\right)}{\partial \xi}=0. \label{moveq}
\end{equation}

By measuring the maximum population density, $u_{max}$, of the
front solution or the steady solution in a moving frame, we obtain
a curve that will be useful in interpreting the following results.
As the bubble accelerates in the moving frame, the system is not
allowed to relax into a steady state, see the accelerating curve
of Fig.\ref{acceleratingbubble}. The time involved in the changes
induced by the acceleration of the bubble compete with the
relaxation time of the system. We will leave these results for now
and recall them later to understand the following analysis.

\section{Oscillating Bubble}

Here, the habitable bubble of fixed width oscillates as a whole.
We study two cases: sinusoidal velocity, and a velocity which is
constant but suddenly changes direction at a given distance away
from the original position. All points within the amplitude of
oscillation will thus be covered periodically for varying amounts
of time. For the purpose of comparison, we will consider the
velocity to be an average velocity for the sinusoidal case. The
parameters involved are the average velocity of the bubble, $v$,
the amplitude of oscillation characterized by the distance between
the center of the bubble at the turning points, $A$, the length of
the bubble, $W$, and the Fisher parameters: the diffusion
coefficient, $D$, the growth rate, $a$, and the competition term,
$b(x,t)$. For the following analysis, $b(x,t)$ is a step function
that defines the shape of the bubble. We have examined the
following aspects of Eq. (\ref{originaleq}):

\begin{itemize}

\item The role of each of the three parameters, $v$, $A$, and $W$, in
the dynamics of the population.

\item How the two cases (sinusoidal velocity and constant velocity)
differ from each other.

\item Qualitative characteristics and comparisons between solutions
for various parameter sets.

\item The evolution of the population as diffusion goes to zero.

\item The appearance of a curious dipping phenomenon.

\end{itemize}

\subsection{Numerical Results for D>0}

It is helpful to divide the behavior into two regimes which
correspond to no-extinction and extinction. The parameters $A$,
$W$, and the critical bubble width, $W_c=\pi \sqrt(D/a)$
\cite{kk,skellam}, are sufficient to define these regimes. In the
steady state, bubble widths smaller than the critical value are
unable to support a population density. This feature now manifests
itself in the amount of overlap between the two extremes of an
oscillating bubble. For $A<W$ there will be an area that is
permanently covered by the bubble.  When this covered area, from
now on referred to as the $overlap$, is greater than the critical
width, the population will always survive to some degree. This
regime is bounded by $0<A<(W-W_c)$.

In the other regime, extinction may occur when there is an overlap
that is less than the critical value, $(W-W_c)<A$. The distance
between the inner edges of the bubble when $A>W$ will be called
the $separation$, $(A-W)$. Naturally, when there is a separation,
no single location will receive continuous coverage by the bubble.
\begin{figure}
\centering \resizebox{\columnwidth}{!}{\includegraphics{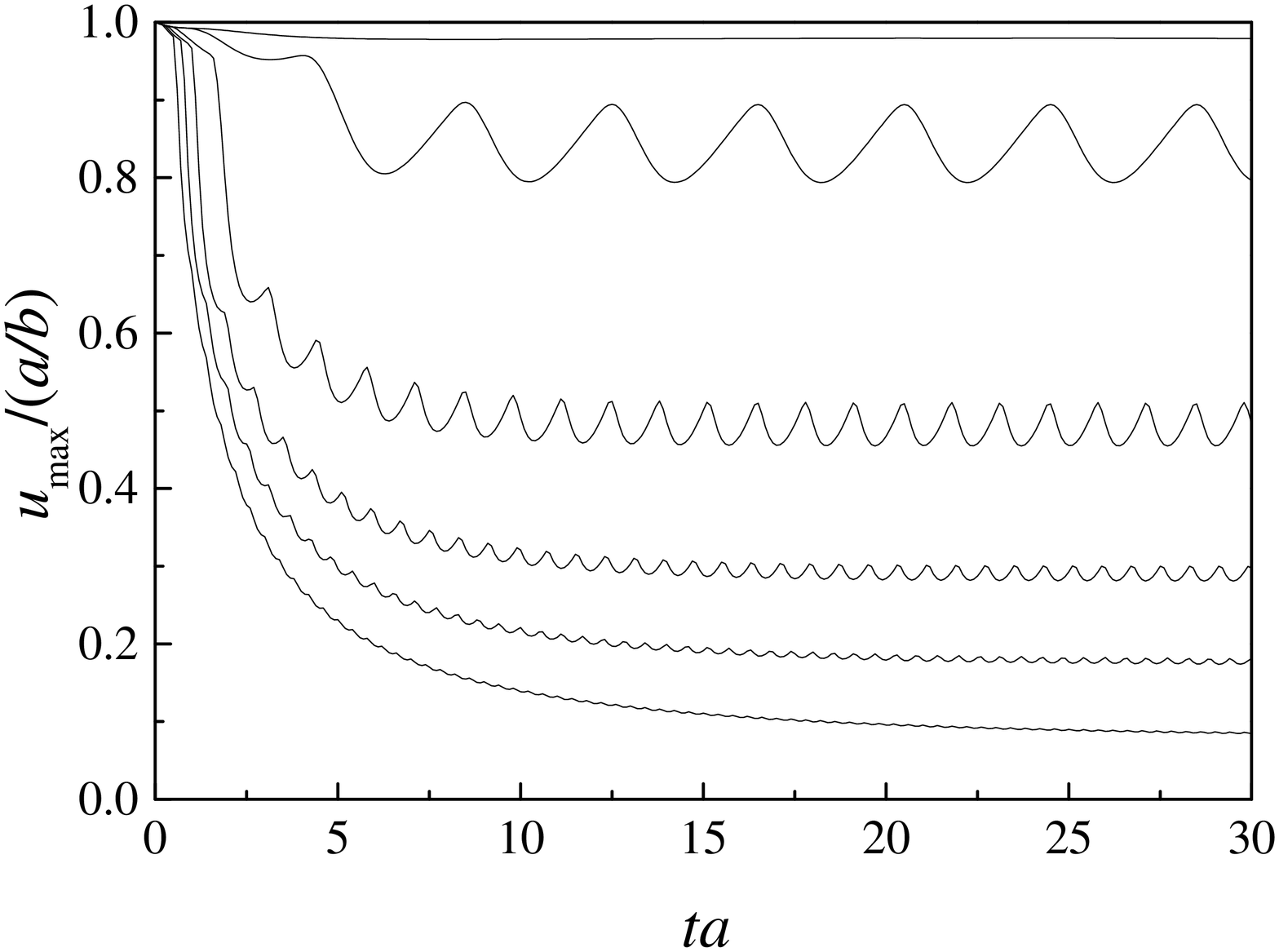}}
\resizebox{\columnwidth}{!}{\includegraphics{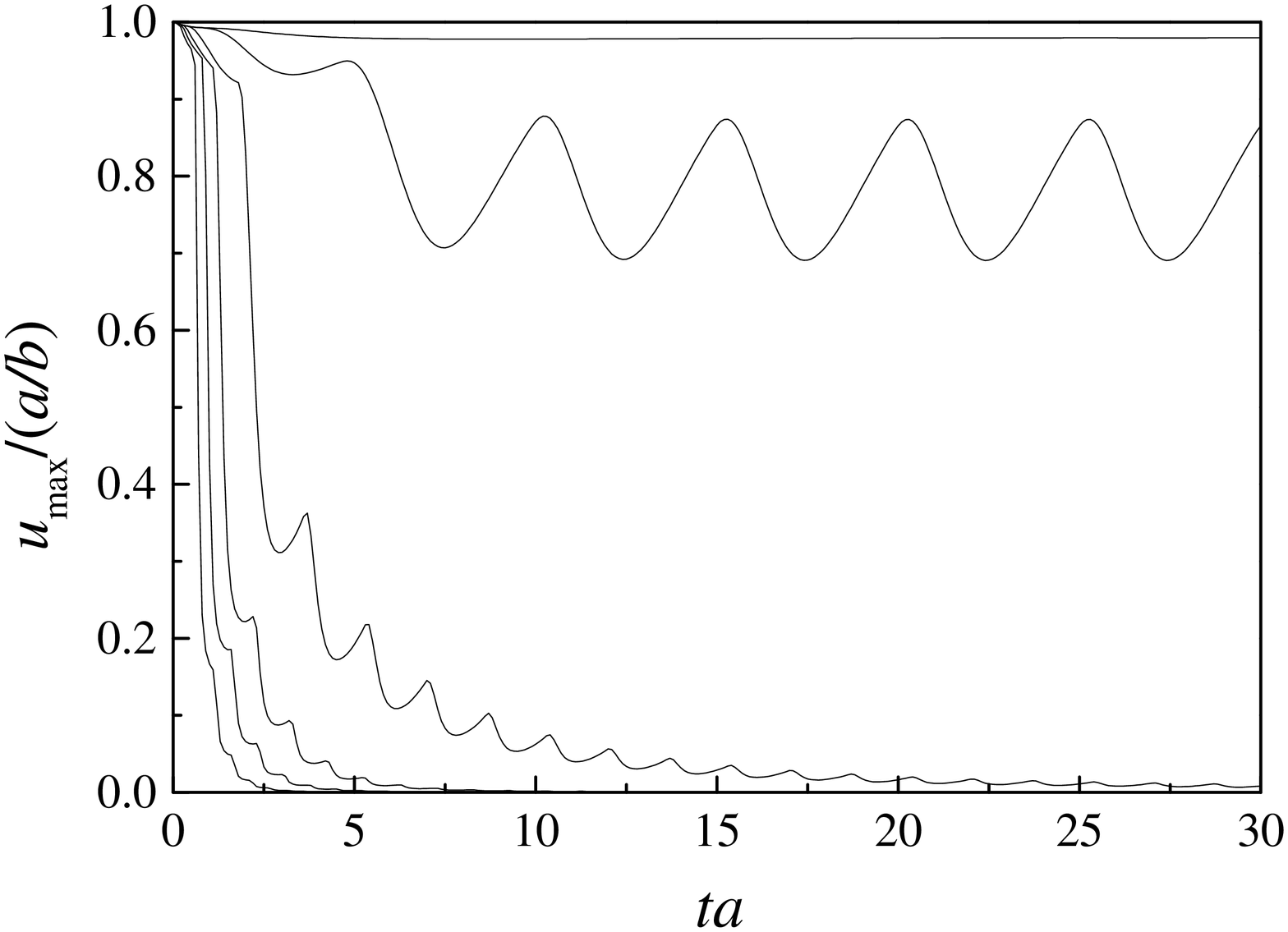}}
\caption{Temporal evolution of $u_{max}$ where in (a) $A=8W_c$,
(b) $A=10W_c$, and $W=10W_c$ in both. Each curve is characterized
by one of seven sinusoidal bubble velocities, $.1aW_c$, $1aW_c$,
$3aW_c$, $5aW_c$, $7aW_c$, and $10aW_c$ where $v_f=aW_c$ when
$D=aW_c^2$, as in the plots above. Higher velocities induce lower
maximum population density values and thus drive a population
closer to extinction as the amplitude of oscillation increases.}
\label{Umaxvs.t}
\end{figure}

\begin{figure}
\centering
\resizebox{\columnwidth}{!}{\includegraphics{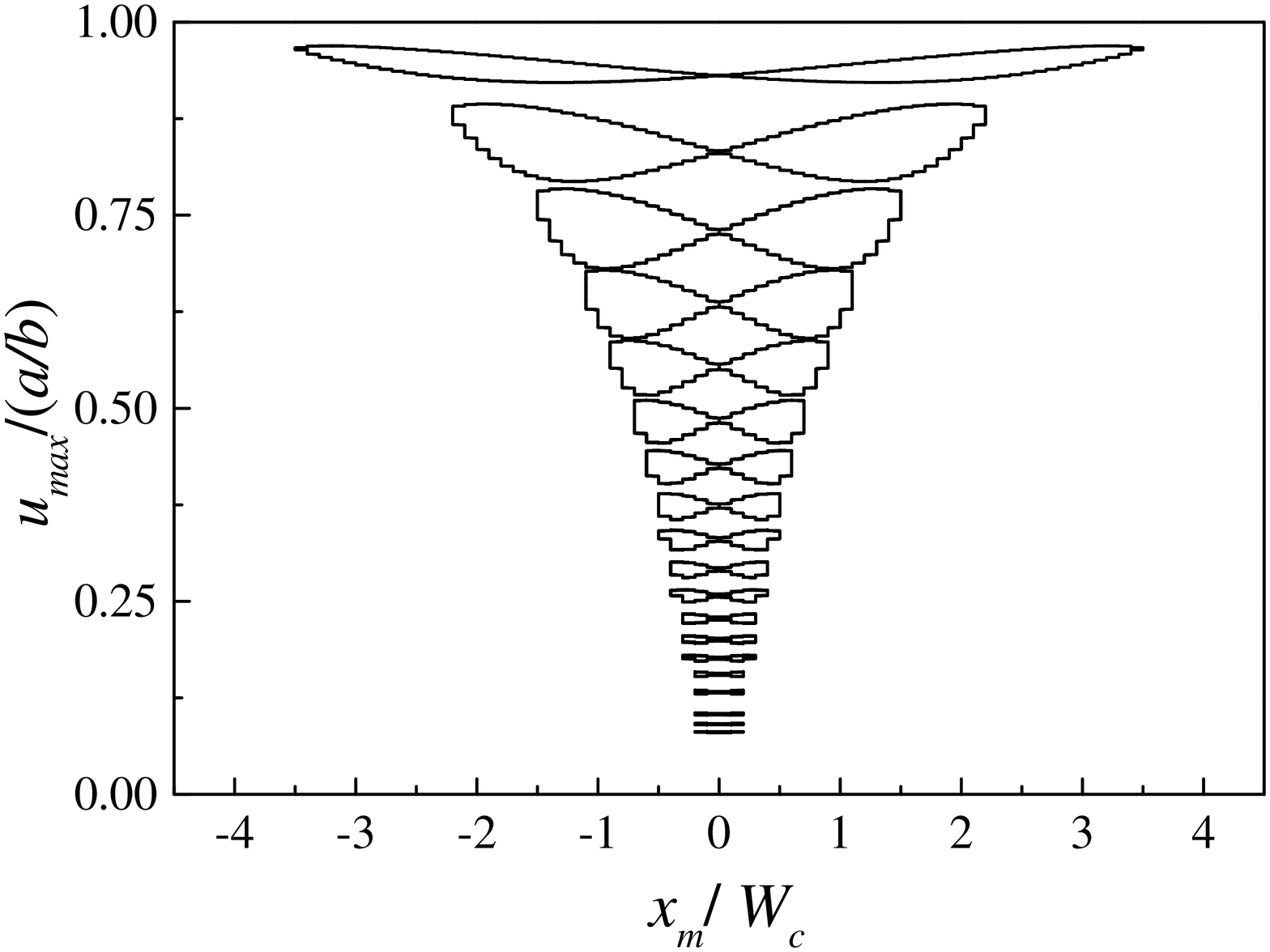}}
\caption{A variety of bowties exhibit the affects for different
values of $v$ when a bubble of $A=8W_c$, $W=10W_c$ and $D=aW_c^2$
moves with sinusoidal velocity. Velocities from top to bottom
range from $aW_c$ to $10aW_c$ and increment by $.5aW_c$ where
$v_f=aW_c$ when $D=aW_c^2$. The trajectories are parametric in
time such that $u_{max}$ is increasing at the endpoints.}
\label{bowties}
\end{figure}

For either regime, if the velocity is below the Fisher velocity,
$v_f$, the population will be able to follow the oscillating
bubble. As the velocity increases it becomes more difficult for
the population density to keep up with the moving bubble (via
diffusion and new growth). Some of the population falls prey to
the harsh environment which quickly reduces the density. If
$b(x,t)$ is infinite for |x|>W/2, the exposed population will go
extinct, never to return. For realistic as well as numerical
purposes, we will consider $b$ outside the bubble to be a large
but finite number. Once outside of the bubble, the population
decreases immensely yet it now retains the ability to regenerate.
Due to the extended amount of time that either end of the bubble's
trajectory is sheltered, peaks of increased population density
form at these extremes.  As the velocity is increased, these peaks
become more pronounced with respect to the maximum population
density in the center of the bubble's trajectory. They also come
more frequently since the bubble traverses the same period of
oscillation more often, see in Fig. \ref{Umaxvs.t} and Fig.
\ref{bowties}. The entire density profile oscillates
spatio-temporally as the velocity increases and varies with the
amount of overlap or separation.

For the zero extinction regime, high velocities effectively wash
out the oscillatory behavior and the peaks merge so that the
length of permanent overlap then determines the population
density. The behavior of the population in the extinction regime,
$(W-W_c)<A$, is the same except for one crucial characteristic.
Again, the peaks at the endpoints come closer together for higher
velocities, but they never merge.  The population dies out before
a constant density forms in the center. As $A$ increases,
extinction occurs for smaller and smaller velocities.

In the discussion above an $average$ velocity of the bubble was
assumed. We now make a distinction between two types of velocities
and their relative affects on the population density. A bubble
moving with sinusoidal velocity lingers for longer time over the
end points of its trajectory and for shorter time in the center as
compared to the constant velocity bubble. For a single location, a
larger population density is permitted to grow if more time is
spent under the bubble. The basic qualitative differences of the
population density thus follow from these two aspects. Though the
two velocities produce the same essential behavior, we find that
there indeed exists a difference between the two after long times.
This will be addressed in the final section.

To further understand the structural changes of the population, we
have examined the maximum population density as it oscillates
parametrically in time. The shape that is outlined can be roughly
characterized by a $bowtie$, Fig. \ref{bowties}. The densities of
the maxima, minima, center, and the spatial extremes are the four
features that comprise the qualitative differences between bowtie
morphologies.
As the average velocity increases these four points of comparison
within the bowtie undergo changes in proportion. The maxima of the
bowtie move closer to the center (recall that in the no-extinction
regime these maxima eventually merge). The minima move farther
from the center. The spatial spread of the bowtie is smaller and
the overall values of the population density decrease for all
positions. The difference between the maxima and minima of the
bowtie, $A_b$ (the amplitude of oscillation of the maximum
population density) initially increases and then decreases as
velocity increases. Fig. \ref{velocityvariance} depicts the abrupt
appearance and then gradual decay of the bowtie as a function of
velocity. The bowtie appears in both nondiffusive systems and
diffusive systems. Bowties of the former develop amplitudes larger
than those of the latter. Also, bowties are formed at lower
velocities for the nondiffusive than for the diffusive system.

\subsection{Analytic Results for D=0}

We begin by considering the Fisher equation where the oscillation
of the bubble is explicitly built into the argument of the
competition parameter $b(x,t)$ of Eq.(\ref{originaleq}).
\[\frac{\partial u(z,t)}{\partial t}
= a\, u(z,t) - b(z+\int_0^{t}v(s)\,ds)\, u(z,t)^2 + D\,
\frac{\partial^2 u(z,t)}{\partial z^2}\]

The transformation $\phi = 1/u$ yields an equation of the form
\begin{eqnarray}
    \frac{\partial \phi(z,t)}{\partial t}
    + a\, \phi(z,t) & = & \nonumber\\
        & &\hspace{-100pt} b\left(z+\int_0^{t}v(s)\,ds\right)
        + D\left[\frac{\partial^2 \phi(z,t)}{\partial z^2} -
        \frac{2}{\phi}\left(\frac{\partial \phi}{\partial z}\right)^2\right].
\end{eqnarray}

By allowing $D=0$, we recover the equation for a {\it highly
damped linear oscillator} that is driven by an external force
proportional to $b(z+\int_0^{t}v(s)\,ds)$. The population will
thus follow the periodic forcing function with a lag. This
oscillator interpretation clarifies the structure of the bowtie.
Each location grows periodically, with a temporal lag, at the
behest of the driving term. The locations at the edges of the
bubble's spatial trajectory will have higher population densities
because the driving oscillator lingers for longer time at a lower
value (less harsh conditions).

An additional transformation
\[x = z + \int_0^{t}v(s)\, ds,\]
and letting $D=0$ leads to,
\begin{equation}\frac{\partial \phi(x,t)}{\partial t} + v(t)\,\frac{\partial
\phi(x,t)}{\partial x} + a\, \phi(x,t) = b(x). \label{nodif}
\end{equation}

\begin{figure}
\centering
\resizebox{\columnwidth}{!}{\includegraphics{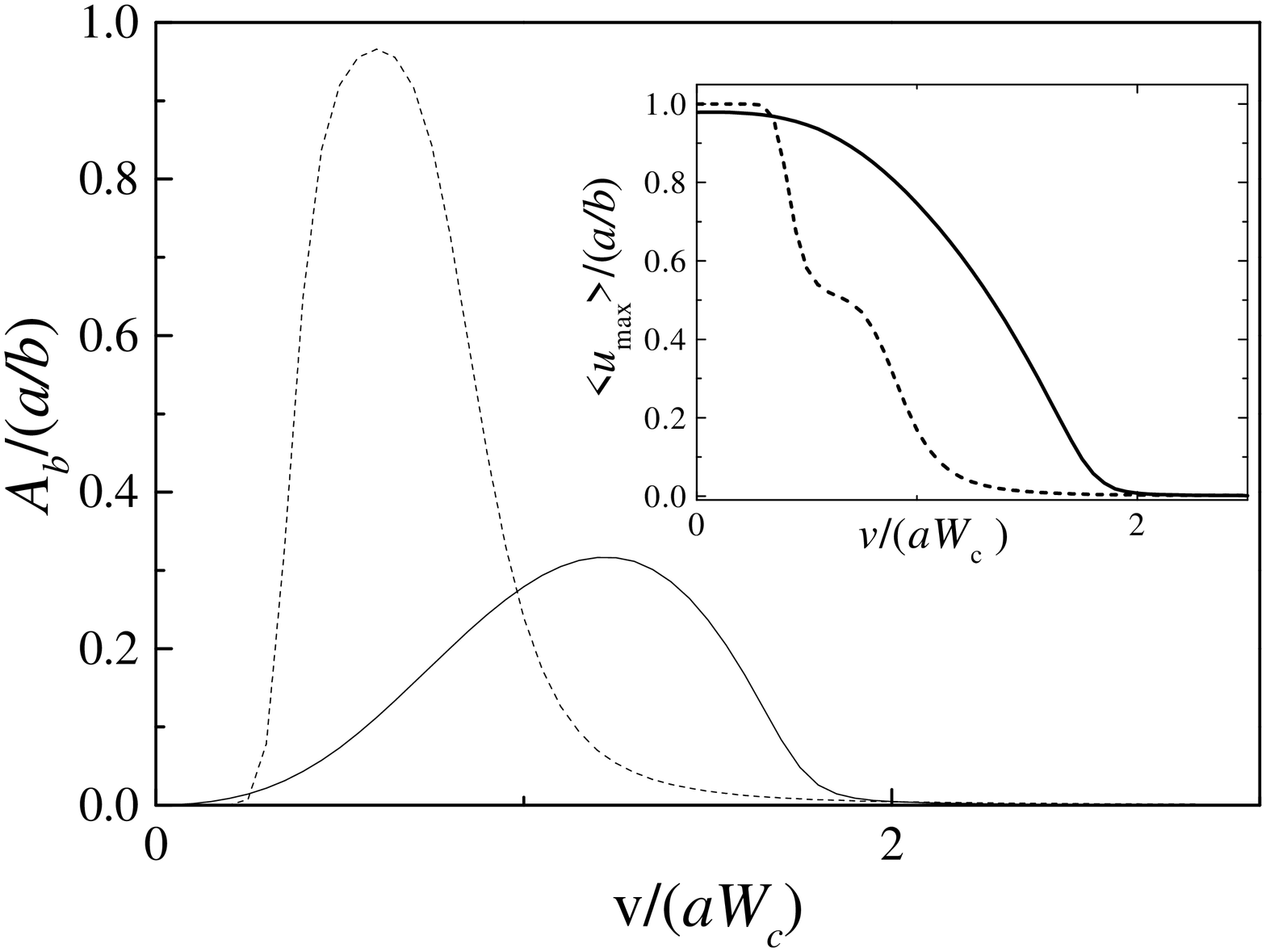}}
\caption{The amplitude of the bowtie is plotted against the
average velocity (which here varies sinusoidally). The inset shows
the average $u_{max}$ after long time. The combination of the both
show the validity of the analytic solution in the high velocity
regime when the curves for $D=aW_c^2$ and $D=0$ converge.}
\label{velocityvariance}
\end{figure}

The solution, as derived in \cite{luca}, was found to be,

\begin{eqnarray}
    \phi(x,t) & = & e^{-a\,t}\,\phi_0\,(x-\int_0^{t}v(s)\,ds)\nonumber\\
        & & \hspace{-30pt} + \int_0^{t}e^{-a\,(t-t^{\prime})}\,
        b\left(x-\int_0^{t^{\prime}}v(s)\,ds\right)\,dt^{\prime}
\end{eqnarray}
and finally, $u(x,t)$ explicitly as
\begin{eqnarray}
    u(x,t) & = & \\
        & & \hspace{-50pt} \frac{u_0(x-\int_0^{t}v(s)\,ds)}{e^{-a\,t} +
u_0(x-\int_0^{t}v(s)\,ds)\,
        \int_0^{t}e^{-a\,(t-t^{\prime})}\,b(x-
\int_0^{t^{\prime}}v(s)\,ds)\,dt^{\prime}}. \nonumber
\end{eqnarray}
As the bubble's velocity greatly surpasses the Fisher velocity,
the effect of diffusion becomes negligible. The diffusion length
per time competes with the velocity of the bubble. Diffusion is
thus limited by the time a single location is covered by the
bubble and has little effect when velocity is high. A solution
thus exists in the high velocity limit, refer to Fig.
\ref{velocityvariance}.

When the velocity is relatively low, $v<v_f$, population migration
occurs via diffusion and new growth. By letting $D\rightarrow0$
(correspondingly, $v_f\rightarrow0$ and critical
overlap$\rightarrow 0$) one might conclude that a population
cannot maintain itself unless there is a bit of overlap. This is
true when the velocity is such that growth cannot occur within the
short period of coverage by the bubble. However, when these two
rates (bubble velocity and $aW$) are comparable, oscillations in
the population density again arise in the diffusionless case. The
bowties in these circumstances vaguely resemble those of before.
When $A<W$ the maximum population density is always saturated;
there are no means by which to leave a sheltered region without
diffusion. When $A>W$, the bowtie limits to a $U$ shape.

\section{Breathing bubble}
\begin{figure}
\centering \resizebox{\columnwidth}{!}{\includegraphics{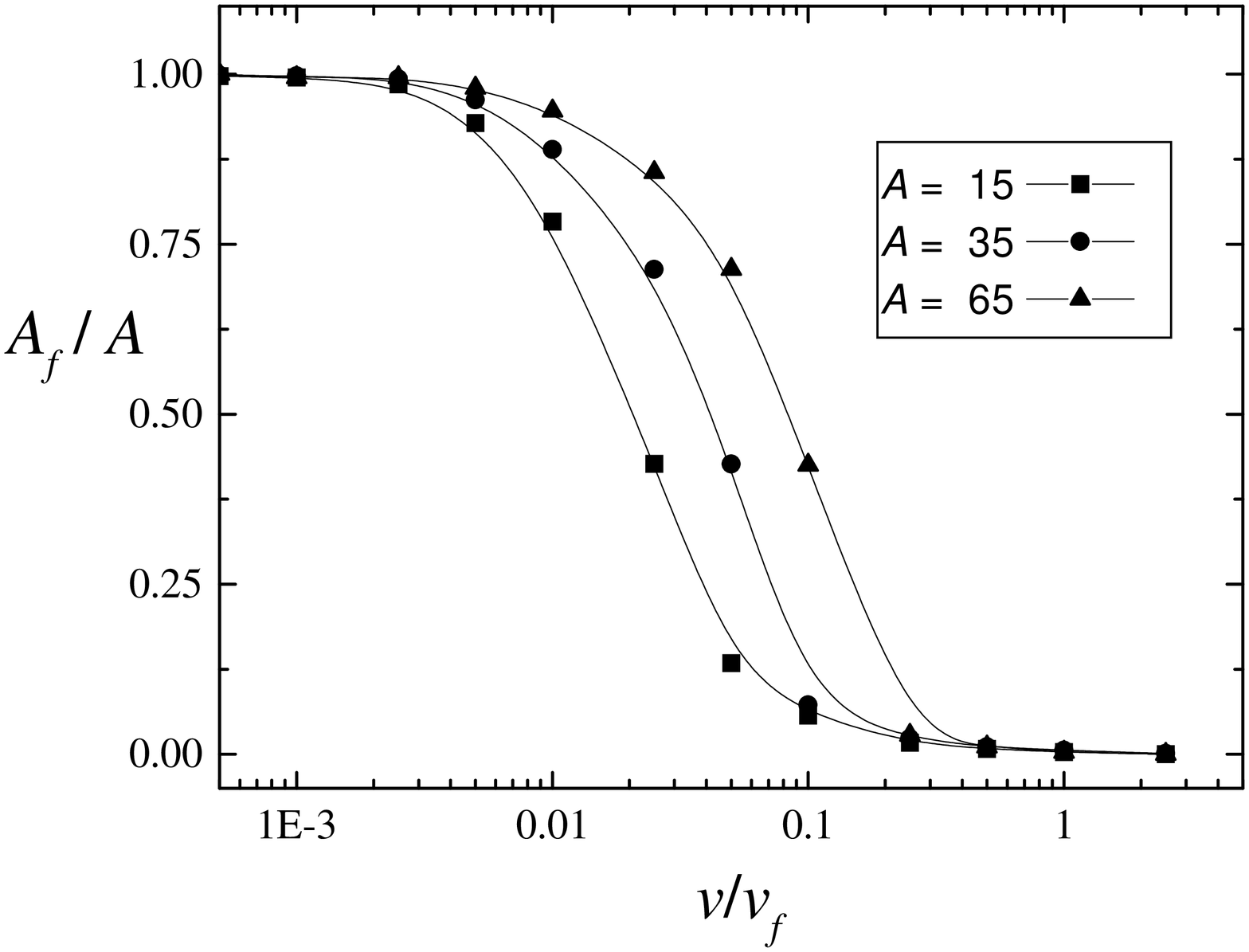}}
\caption{Amplitude of the front oscillations for several
velocities of the bubble where $D=aW_c^2$.} \label{fig1}
\end{figure}

The breathing bubble corresponds to a situation when the width of
a anchored bubble varies in time. As before, we consider several
cases and for each of them a family of values for the involved
parameters. These parameters are the velocity of the variation of
the width, the mean width of the bubble, the amplitude of the
breath, and again, the Fisher parameters of Eq.
(\ref{originaleq}). We are interested in analyzing the
relationship between the existence of critical values and the
response of the population density to a changing environment. We
first take the case when the bubble width oscillates at a constant
speed but always preserves a size above the critical one. The
interesting aspect of this situation occurs at the boundaries of
the bubble where population fronts are formed. The fronts can
follow the movement of the bubble when the velocity of the
breathing motion is under a critical value. Above this critical
value, the population is confined to the minimum size of the
bubble, which indicates that the population front can no longer
follow the changes of the bubble. To show this we display in Fig.
\ref{fig1} the relative amplitude of the oscillation of the
population front, $A_f/A$ vs. the speed of the bubble breath,
where $A$ is the amplitude of breathing.
We observe that at a critical value this amplitude goes to zero.
That is, in spite of the changing size of the bubble, the
population density remains bounded statically.

Instead of a constant value for the velocity of breathing, we can
take a fixed frequency for a sinusoidal behavior. The frequency is
associated with the breathing motion of the bubble boundaries. The
results are analogous to those displayed in Fig. \ref{fig1}.
\begin{figure}
\centering \resizebox{\columnwidth}{!}{\includegraphics{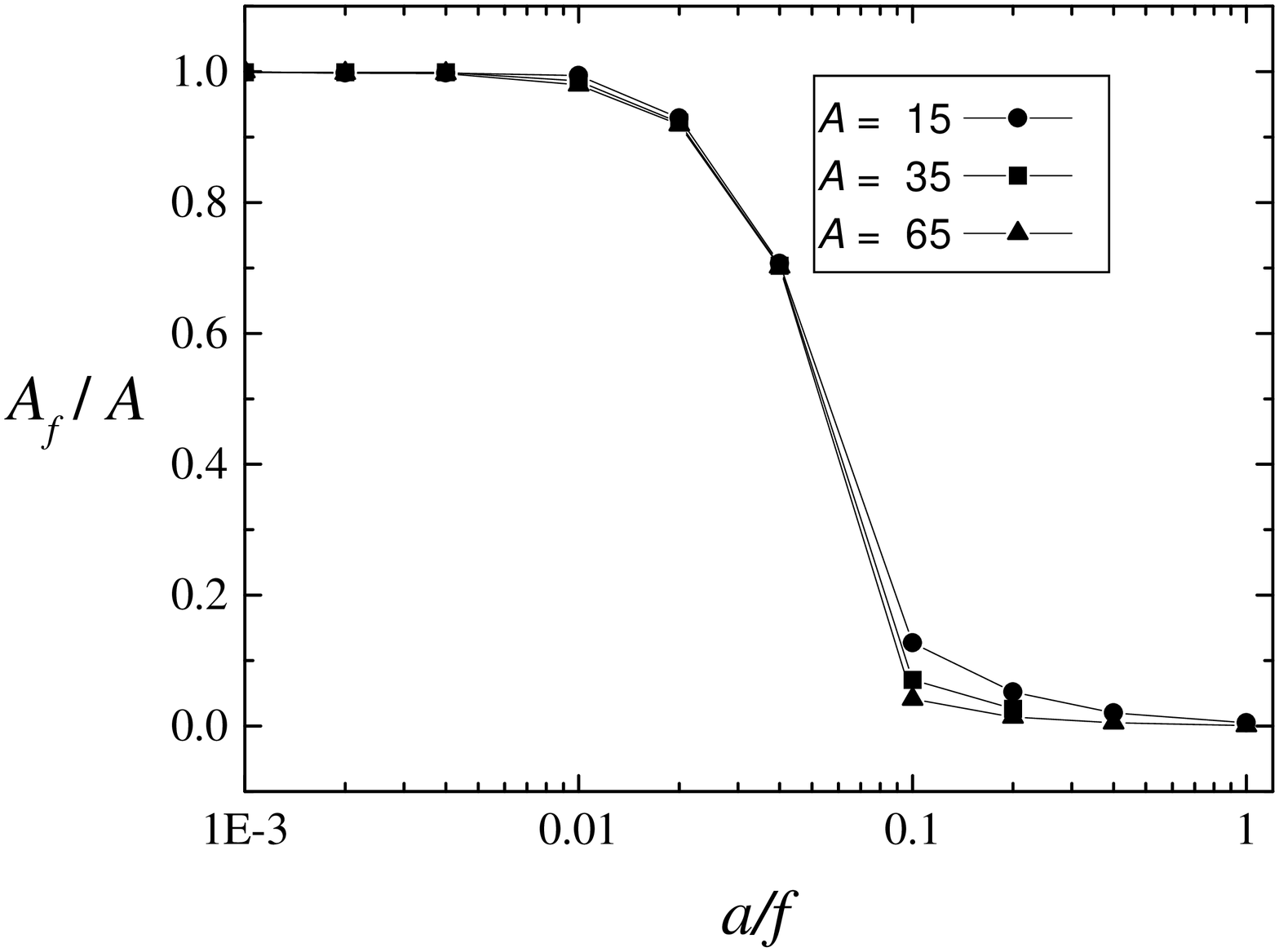}}
\caption{Amplitude of the front oscillations for several values of
the inverse of the frequency of breathing of the bubble where
$D=aW_c^2$.} \label{fig2}
\end{figure}
No qualitative differences were found.

\begin{figure}
\centering \resizebox{\columnwidth}{!}{\includegraphics{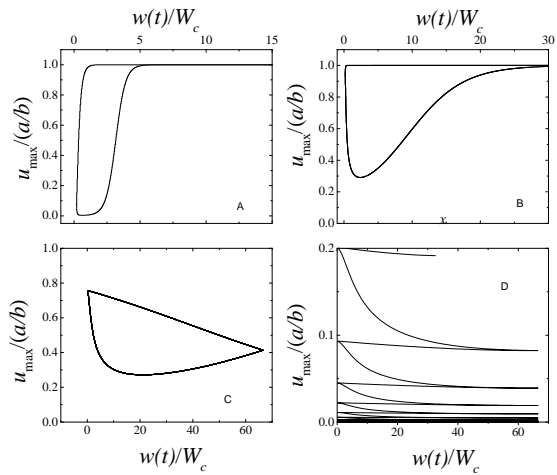}}
\caption{Maximum value adopted by the population as a function of
the instantaneous width of the bubble. The velocity of breathing
is: a)0.1$aW_c$ b)1$aW_c$ c)10$aW_c$ d)40$aW_c$, and $D=aW_c^2$.}
\label{fig3}
\end{figure}

When the minimum size attained by the bubble during the breathing
is under the critical size, the situation is different. It is not
enough to plot, as in previous examples, the amplitude of the
oscillation of the front because some variations in the maximum
density of the population are also observed. To visually
understand what is happening we again plot the temporal behaviour
of maximum value of the population, $u_{max}$ as a function of the
momentary width of the breathing bubble. This is what is shown in
Fig. \ref{fig3}, where the trajectories are traversed clockwise.
As observed in the plots, if the velocity is low enough, the size
of the bubble will be under the critical size long enough to
provoke the extinction of the population. On the contrary, for
higher values of the velocity there is a range of values at which
the population can survive. The response of the population to
changes is much slower than the dynamics of the environment. Each
of the results obtained in this case are essentially the same as
the corresponding results from the oscillating bubble. In
particular, the results displayed in Figs. \ref{fig1} and
\ref{fig2} can be associated with the discussion included in the
previous section, those included in Fig. \ref{fig3} are analogous
to the bowtie effect.

\section{Further features}
\begin{figure}
\centering
\resizebox{\columnwidth}{!}{\includegraphics{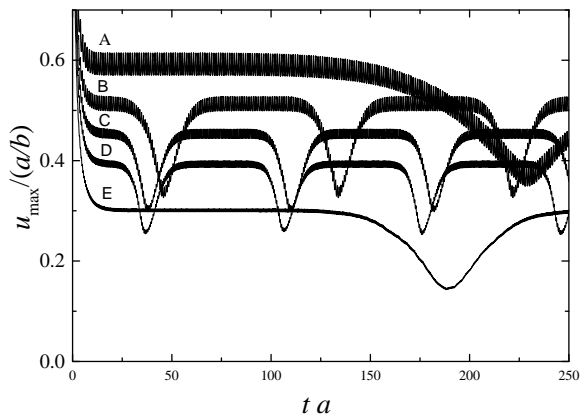}}
\caption{After long times a periodic dip in all values of $u_max$
is observed for constant velocities. The sampling above, (A, B, C,
D, and E) corresponds to velocities (3.5$aW$, 4.5$aW$, 5.5$aW$,
6.5$aW$, and 8.5$aW$, respectively) where $v_f=aW$ because
$D=aW_c^2$.} \label{phenomena}
\end{figure}

We now return to a surprising feature that arises only in the
presence of constant bubble velocity and diffusion. Probing longer
times reveals an overlying oscillation in the population density
profile. For diffusive systems, $D>0$, and constant velocities,
the entire bowtie dips down and then back up again in a periodic
manner with constant amplitude. For different velocities the
bowtie dips and rises with different periods and amplitudes, see
Fig. \ref{phenomena}. Also, notice that as velocity increases, the
frequency of the alternative dip doesn't simply increase as well.
Nor does the amplitude of the dip appear to change with the
velocity. Sinusoidal velocities do not produce this affect.
Further analysis is necessary to completely understand the origin
of this behavior and its relation to the two required conditions,
diffusion and constant bubble velocity.

\section{Conclusions}

In the present work, we have numerically studied the behavior of a
population whose dynamics are described by a modified Fisher
equation with additional environmental variations. Analytic
solutions were also considered for simplified examples. The
results presented here show that besides the expected behavior
some surprising features arise. There exist two main parameters
that characterize the evolution of a given population: the Fisher
length and Fisher velocity. A given group of individuals will not
survive if the habitat size is smaller than the Fisher length. At
the same time, the population will not be able to maintain
saturation when the habitat moves with a velocity higher than the
Fisher velocity. With these two considerations we have examined
two types of periodically varying environments. Though extinction
of the population was verified in conditions corresponding to the
two situations mentioned above, it was also found that some
nontrivial scenarios can also arise. Situations in which
extinction was expected exhibited what we called the bowtie
effect. An interplay between the time associated to the transient
towards extinction and the periods of the changing environments is
the cause of unexpected behavior of the population. We have also
verified that there is a correspondence between the results
obtained when considering oscillatory or breathing bubbles, though
the situations are different. A diffusionless approximation helped
to understand the origin of the observed phenomena in terms of
oscillating perturbations. Perhaps the most unexpected result is
that depicted in Fig.\ref{phenomena}. An explanation for these
features is still in progress.

\section{Acknowledgements}

This work is supported in part by the Los Alamos National
Laboratory via a grant made to the University of New Mexico
(Consortium of the Americas for Interdisciplinary Science), by the
NSF's Division of Materials Research via grant No DMR0097204, by
the  NSF's International Division via grant No INT-0336343, and by
DARPA-N00014-03-1-0900. 

\end{document}